\documentclass[conference]{IEEEtran}
\ifCLASSINFOpdf
\else
\fi
\usepackage{url}

\usepackage{adjustbox}
\usepackage{array}

\newcolumntype{R}[2]{%
	>{\adjustbox{angle=#1,lap=\width-(#2)}\bgroup}%
	l%
	<{\egroup}%
}

\usepackage{fontawesome}

\usepackage{listings}

\begin{document}
%
\title{The Inconvenient Truths of Ground Truth for Binary Analysis}

\author{\IEEEauthorblockN{Jim Alves-Foss}
\IEEEauthorblockA{Center for Secure and Dependable Systems\\
University of Idaho\\
jimaf@uidaho.edu}
\and
\IEEEauthorblockN{Varsha Venugopal}
\IEEEauthorblockA{Center for Secure and Dependable Systems\\
	University of Idaho\\
	vvenugopal@uidaho.edu}
}
	

%


\IEEEoverridecommandlockouts
\makeatletter\def\@IEEEpubidpullup{6.5\baselineskip}\makeatother
\IEEEpubid{\parbox{\columnwidth}{
		Workshop on Binary Analysis Research (BAR) 2022 \\
		24 April 2022, San Diego, CA, USA \\
		ISBN 1-891562-76-2 \\
		https://dx.doi.org/10.14722/bar.2022.23010 \\
		www.ndss-symposium.org
	}
	\hspace{\columnsep}\makebox[\columnwidth]{}}

\maketitle

\begin{abstract}
The effectiveness of binary analysis tools and techniques is often measured with respect to how well they map to a \textit{ground truth}. We have found that not all ground truths are created equal. This paper challenges the binary analysis community to take a long look at the concept of ground truth, to ensure that we are in agreement with definition(s) of ground truth, so that we can be confident in the evaluation of tools and techniques. This becomes even more important as we move to trained machine learning models, which are only as useful as the validity of the ground truth in the training.

\end{abstract}


%

\section{Introduction}
Binary analysis research involves automatic analysis of executable binaries and a transformation of those binaries into some intermediate representation that allows for complex analysis. Part of this transformation involves the disassembly of the binary into low level constituent pieces: instructions and data, and also into more complex constructs: functions, classes, data structures. There have been many research projects and tools developed to assist with this process. The question is how good are they?  To determine the effectiveness of the tools requires an understanding of the correct answers and then how well the tools map to those answers. These correct answers are the {\em ground truth} of the system.

There are many ways that researchers find the ground truth of the test suites they are using to evaluate their tools, and to generate results for publication. Unfortunately, this part of their work is often not seen as interesting and so the details of ground truth generation are not published in their papers, and scripts or directions for ground truth generation are not made available in the public repositories for the tools. Some readers may respond ``but that is obvious'' or ``that part is easy''. Although we have heard these types of responses at meetings, the devil is in the details, and what may initially appear easy may not be the real ground truth. For example, the Byteweight project \cite{Bao14b,Bao14} published results related to detecting function boundaries in binary executables. Their software returned start address and number of bytes in a function. Their script that was used to evaluate the correctness of the reported length actually reported any short lengths as correct. This was not discussed anywhere in the publication or documentation of the tools. We assume that this was a {\it hack} since the symbol tables of unstripped binaries generated by the Intel compiler icc include padding bytes as bytes of the preceding function, while the gcc compiler does not include these. Two different compilers generated two views of the ground truth. We would like to avoid these hacks in the future. In addition, there were instances of multiple symbols referencing the same function in some binaries. The analysis counted those functions twice. In a recent study, Koo et al.~\cite{Koo22} generated a set of test suites to evaluate the function identification problem. A couple of the binaries did not have full symbol table information and therefore the ``ground truth'' created for the study was initially incorrect. They examined this issue and adjusted their data, and explained it in their paper. This problem could have been easily missed in other studies. 

We argue that there needs to be research into, and consensus on the evaluation/testing tools and the generation of test suites. Therefore when researchers evaluate the results of a new tool or technique, the community agrees with the measurements of the evaluation. In the general sense, we are talking about applying the science of instrumentation not only to our binary analysis tools, but to the tools that generate the ground truth values that we use to evaluate the binary analysis tools. 

The purpose of this paper is to discuss the concepts of ground truth, to evaluate what has been published related to ground truth, and to start a dialog with the community to develop a set of standards for consistent ground truth determination. We would also like to see easily accessible and publicly available tools for ground truth generation and reporting.

We examined over 50 binary analysis papers, and found that many of them did not discuss how they calculated ground truth. Those that did, either used the generated symbol table, generated debug data, another tool such as IDA Pro, modified the compiler, or manually determined the ground truth.  This paper summarizes the main categories of ground truth we found, the techniques to extract ground truth and possible confusion from this techniques. 

Before we start, we want to mention that there have been other papers that touch on parts of this topic. One notable paper is by Pang et~al.~\cite{Pang21} that evaluates nine binary analysis tools/platforms. They instrument both clang and gcc compilers to determine the ground truth for disassembly, symbolization (references to other code/data), function entries, and control flow graphs (indirect jumps and calls, tail calls and non-returning functions). Li~et~al.~\cite{Li2020} grab intermediate information from the compiler to generate their disassembly ground truth, in an attempt to validate ground truth. Another useful paper is by Kim et~al.~\cite{Kim21} that discusses a benchmark for binary code similarity analysis.

\section{Defining Ground Truth}

The definition of {\em ground truth}, with respect to a binary executable, is context dependent. A lot depends on what the researchers are examining: correct decoding of instructions and data, mapping of binary to complex structures such as functions or data structures, or mapping of binary back to the source code. This context matters when evaluating the effectiveness of binary analysis tools. It also matters if we are examining normal code, obfuscated code or malware. For the purposes of this paper, we will primarily focus on normal code. 

Compilers perform optimizations, and may add or delete portions of the code, so that there is not an one-to-one mapping between source code and binary. Most compilers insert standard functions for code initialization and termination, and other may insert new optimized libraries (such as Intel compiler's embedded memory management routines \cite{icc}). 

Therefore, when researchers are analyzing a binary in the context of what the binary does, we argue that they must look at the binary as compiled. In the context of functions, researchers must first define what a function is, within the context of the binary, as compiled and linked. For example if a source code function is in-lined, it is no longer a function, and an analysis tool should not claim it found that function within the binary. A more difficult question we have to ask, is if the compiler optimizes out tail calls\footnote{A tail call exists when the final instruction of a function is a call to another function. An optimizing compiler will clean up the local function context, and then jump to the next function instead of using a call, saving the expense of a call and return.}, is a function that is only jumped to, instead of called, still a function?

\subsection{Instructions}
The first category we examine is instructions versus data for disassembly of a binary. It is well known that it is hard to analyze a binary if the tool does not correctly identify instructions and data within the binary. Compilers, or hand written assembly may embed data within code sections of the binary, making it difficult to have precise determination of the mapping of bytes from the binary to instructions. Questions researchers should ask are:

\begin{itemize}
\item{\em Where do we obtain the ground truth about which bytes are instructions?} Do we need to instrument the compiler, or is their sufficient details in the debug data?
\item{\em What about instructions as data?} There are some programs that run checksums over their own code, in this case code is data -- most people will define that code as instructions. However, other programs may store portions of code that they copy to other parts of memory. This may be malicious, or may be part of a just-in-time compilation routine. Is that stored code really instructions, or just data?
\end{itemize}

{\bf Recommendation:} We recommend defining instructions as executable code as compiled into the binary. Instructions that are  only meant to be copied, and are therefore used only as data, should be categorized based on location. If these are stored in a data section, then they are data. If these instructions are stored in a code section, they are code, even if they are never called or executed.

\begin{lstlisting}[caption={Multiple entry points from binary {\tt as\_new} (binutils v 2.23). Compiled with icc v 14.0.1, in 32-bit mode, -O2 option. Shown using AT\&T syntax.}, captionpos=b,label={lst:two-entry},frame=tb,float,floatplacement=T]
fix_syms(): ../binutils2.23/bfd/linker.c:3208
080b41c0 <fix_syms>:
 80b41c0:	mov    0x4(%esp),%eax
 80b41c4:	mov    0x8(%esp),%edx

080b41c8 <fix_syms.>:
 80b41c8:	push   %esi
 80b41c9:	push   %edi
\end{lstlisting}

\subsection{Functions}
Let's take functions as the next example of a construct for which a researcher may want to find the ground truth. At a high level of abstraction, a function is a construct in a high level programming language. We can analyze source code and clearly find function names, types, parameters and boundaries. If a researcher is looking at binary differencing \cite{BinDiff}, traceability, or some other concept where there is a need to map the source code directly to its implementation in a binary, then the ground truth we are looking for must include that mapping. 

However, a compiler can apply many optimizations, and not directly map the source code to the executable code. For example, functions may be in-lined, where there is no actual function call within the binary, but rather the body of the function is directly compiled into the body of the calling function, effectively treating the function as a macro. There may be optimizations involving tail calls. In addition, compilers may optimize local variables so that they are stored in registers only, or do loop unrolling, or many other optimizations that prevent a direct mapping between source code and the binary. Therefore, we believe researchers must address the following questions:

\begin{itemize}
\item {\em Is there a need to map binary directly to source code?} If so, what happens when a function is in-lined? What happens if the compiler creates multiple entry points to the function, or divides the function into two separate functions?
\item {\em What is a function within the binary?} Do we define a function with a strict adherence to the source code? What about a function with a tail call$^1$? If the compiler optimizes away the call with a jump, are we jumping to a new function or to a disjoint portion of the same function? What about a compiler that allows a function to fall through to the next ``function''?  
\item {\em Are uncalled functions still functions?} Some compilers may optimize them away, while others do not. 
\item {\em Are compiler added functions, functions?} There are several functions added by compilers, many are hand coded assembly.
\item {\em Can functions have more than one entry point?} Some compilers may insert multiple entry points for the same function. The Intel compiler does this in 32-bit mode (see discussion in Section~\ref{two-entry}, and see Listing~\ref{lst:two-entry}). In addition, some compilers will optimize a function into multiple separate functions (See discussion of Listing~\ref{lst:as-new-syms} in Section~\ref{multiple-functions}.)  
\item {\em Besides function entry points, how do we define function boundaries?} Some tools look for the start and end bytes of a function. Is the listed end byte the last byte of the last instruction? The first byte of the last instruction (for multibyte instructions)? The first byte after the last instruction? And how about padding between functions, does that belong to a specific function? What if the last instruction is unreachable?

\end{itemize}

{\bf Recommendation:} We recommend defining functions as logical units within the binary that have an entry point and one or more exit points and a collection of instructions in between. The exit may be a call to a non-returning function, a jump to the start of another function, or a return. This maps as closely as we can to the source code. We do not count in-line functions as functions, but rather treat them as if they were macros. We do count compiler inserted functions as functions, since they are in the binary. The hard part is determining the end of the function. Ideally we want to define a function as all of the bytes from the start of the function until the start of the next function, assuming functions are contiguous.

\subsection{Function parameters and Local variables}
It is possible that an optimizing compiler may remove or add parameters to a function, although we have only seen this in experimental compilers. As we mention in Section~\ref{multiple-functions}, some compilers can optimize the values of constant parameters and the ignore the passed in value. Compilers can also optimize away local variables, keeping the values only in a register. Therefore, we believe researchers must address the following questions:

\begin{itemize}
\item {\em How are you defining parameters?} There is a need to map parameters back to source code, to registers or stack locations where they are stored, or maybe for data flow graphs. If a compiler optimizes away the use of a parameter, what is the ground truth in the mapping to source code, or to data flow graphs?
\item {\em How are you defining local variables?} There is also need to map local variables back to source code, to registers or stack locations where they are stored, or maybe for data flow graphs. If a compiler optimizes away the use of a local variable what is the ground truth in the mapping to source code, or to data flow graphs? Is there a need to understand the location of the local variables on the stack -- since some compilers will change the order.
\end{itemize}

{\bf Recommendation:} We recommend defining parameters and local variables as those that are compiled into the binary. If the parameter or local variable is optimized away, it does not exist in the ground truth of the binary. 

\subsection{Cross references/pointers, Indirect jumps and calls}
Within a code section there will be pointers/references to other code and data. Within a data section, there will be pointers/references to code and data. These pointers/references may be absolute values, or relative. We don't think there are too many questions related to these values as they are what they are.

There are jump tables, function pointers and other values that are used for indirect jumps and calls. The location of these values fits into this category as pointers/references to code. The question researchers have to ask here is:

\begin{itemize}
\item{\em Where do we obtain the ground truth about which bytes are these references/pointers?} Some people recommend using relocation data for pointers, but that does not work well in position independent code which does not need as much relocation information. 
\end{itemize}

{\bf Recommendation:} These values are what they are. We don't think instrumenting a compiler to find these values is the best long-term solution for generating the ground truth.If debug data contains all typing information, we can extract it from there. (See Section~\ref{Recommend} for further discussion).

\begin{lstlisting}[caption={Symbol table from binary as\_new (binutils v 2.23). Compiled with icc v 14.0.1, in 32-bit mode, -O2 option.}, captionpos=b,label={lst:as-new-syms},frame=tb,float=*,floatplacement=T]
08055750 l     F .text	00000c30              operand..0
080570a0 l     F .text	00000330              integer_constant..0
08056b60 l     F .text	00000540              integer_constant..2
08056840 l     F .text	00000320              integer_constant..3
08056520 l     F .text	00000320              integer_constant..4
0805a7d0 l     F .text	00000cb0              expr..0
080573d0 l     F .text	00000c80              expr..1
08058fa0 l     F .text	00000cd0              operand
08056380 l     F .text	000001a0              integer_constant..1
\end{lstlisting}

\begin{lstlisting}[caption={Code samples from binary as\_new (binutils v 2.23). Compiled with icc v 14.0.1, in 32-bit mode, -O2 option. Shown using AT\&T syntax.}, captionpos=b,label={lst:as-new-samples},frame=tb,float=*,floatplacement=T]
integer_constant..2 (radix is 16)
/data/usenix/Linux/binutils-2.23/gas/expr.c:360
      number = number * radix + digit;
 8056baa:	8b d5                	mov    %ebp,%edx
 8056bac:	c1 ea 1c             	shr    $0x1c,%edx
 8056baf:	c1 e5 04             	shl    $0x4,%ebp
 8056bb2:	c1 e6 04             	shl    $0x4,%esi
 
integer_constant..3  (radix is 2) 
/data/usenix/Linux/binutils-2.23/gas/expr.c:360
      number = number * radix + digit;
 805688c:	8b d3                	mov    %ebx,%edx
 805688e:	03 db                	add    %ebx,%ebx
 8056890:	c1 ea 1f             	shr    $0x1f,%edx
 8056893:	03 c0                	add    %eax,%eax
\end{lstlisting}

\subsection{Special functions}
Some functions do not return, and these affect the validity of control flow analysis. Also, some functions are inserted into the binary by the compiler, and are not linked to the source code.  The question researchers have to ask here are:

\begin{itemize}
	\item {\em For non-returning functions, where do we get the ground truth about this characteristic?} We have seen tools that embed a list of non-returning standard library functions. This needs to be updated as libraries change, and may be language or even compiler dependent. In addition, use code can be non-returning if it always calls a non-returning function such as an error handling routing that then call exit or abort. How are these documented?
	\item {\em How does a non-returning function affect function analysis or control flow?} Is an instruction after a call to a non-returning function part of the same function or not? We have seen fault-tolerant code that adds additional instructions after calls to non-returning functions, just in case an error results in a return. We have seen compilers that include additional function clean up instructions, because the compiler can not determine it is a non-returning function. The ground truth of the program must capture this. 
	\item {\em How do we handle uncalled functions?} Some of the compiler inserted functions are actually never called. If library is linked in, all of the functions in the library are usually included, not just the called ones. Are these unused, uncalled functions part of the ground truth?
	
\end{itemize}

{\bf Recommendation:} Non-returning functions are special and need to be recognized and documented. OS system calls may need to be documented with the non-returning attribute. Everything should be recursively analyzable. Any functions inserted by the compiler or linker are still part of the binary and need to be treated as functions in the program.

\section{Examples of ground truth confusion}
Many tools are designed to analyze binaries, \textit{in the wild}. Therefore they may assume that the binaries are stripped of all metadata, including symbol tables and debug data. They may even assume that any symbol table or debug data in the binary may be deliberately incorrect to help thwart reverse engineering. This is a reasonable approach to take. However when we are conducting experimentation to determine how well a tool works, we have control over the experimentation test suite, and we can therefore be confident that any metadata is non-malicious. Researchers should also take care that the test suite is built in a way that allows generation of correct ground truth. For example, Koo et~al.~\cite{Koo22} mentioned that they had some test cases that were missing function data in the symbol table, although it existed in the debug data. We were able to rebuild these binaries with the full symbol table information. If we could not do this, we would recommend removing these outliers from the test data, if ground truth could not be reliably determined.

\subsection{Compiler issues}
Compilers will generate executables where there is not a direct one-to-one mapping from the source code. As an example, the following two examples cause confusion and difficulty when generating ground truth related to function boundaries.

\subsubsection{Compiler insertion of multiple entry points}
\label{two-entry}

Listing~\ref{lst:two-entry} is a code snippet  from the {\tt as-new} binary, from binutils version 2.23, compiled with Intel compiler icc, version 14.0.1 into a 32-bit binary with the -O2 optimization option. Here we see two different entry points for the function {\tt fix\_syms}. The first entry allows for passing of parameters via the stack. These parameters are then stored in registers and then the code falls through to the next function. This allows for linking with other binaries that expect parameter passing on the stack, but leads to ground truth confusion.

What is the ground truth in this case? The code is supporting two different calling conventions, with an optimized register convention being used for calls from functions within the same compilation unit. Is this one function or two? The symbol table says two.

\subsubsection{Compiler duplication/separation of functions}
\label{multiple-functions}

Listing~\ref{lst:as-new-syms}  is a portion of the symbol table from the {\tt as-new} program mentioned in Section~\ref{two-entry}. The first column is the byte address of the symbol, in this case the function start address. The next two columns include flags and a type of the symbol (in the example, all symbols are local and functions). The next column is the name of the section in the binary where the symbol is located, followed by the size -- which is the length of the function.

Here it is evident that the function {\tt integer\_constant} is compiled into 5 separate functions. 
This function takes a string and parses it into an integer, depending on the specified radix (base) of the number. An analysis of the code shows that the first argument of {\tt integer\_constant}, the radix, is passed as a constant in this program. For calls to {\tt integer\_constant..0} it is 10, for {\tt integer\_constant..2} through {\tt integer\_constant..4} it is 16, 2 and 8 respectively. There are no calls to {\tt integer\_constant..1}, where the radix is not one of the standard constants. The code makes very specific choices 
with respect to the value of radix. For each of these functions the compiler optimizes in the constant value of the radix, generating code for just the appropriate subset of the function for that radix. See code snippets in Listing~\ref{lst:as-new-samples}, where the radix is 16 and then 2.

What is the ground truth in this case? If we are mapping source code to binary, we have multiple mappings. If we are looking at individual functions in the binary, there are five functions for this one source function. If we are looking at parameters, we see that the first parameter is optimized out for four of the functions, and is not used, although it is still passed on the stack.

\subsection{IDA Pro}
IDA Pro \cite{ida} is a binary analysis tool that is used by a large portion of the community. It is a good tool with a lot of capabilities and features. However, it is still a tool, and is not necessarily the oracle of ground truth. We have seen papers and projects that say they use IDA Pro for their ground truth, such as \cite{Karamitis18}. This is the most concerning when they use it for ground truth that is then used for training a machine learning algorithm. The results are then compared to IDA Pro generated ground truth. So, although the experimental results may be favorable to the researchers, it does not mean that the trained and evaluated ground truth values are actually correct. The following are concrete examples of this.

For our research, we have compiled several different tool suites using several different compilers and compiler optimization flags. 
We compiled objdump, from the standard binutils package, using the clang compiler with -O2 optimization option for 32-bit Intel architecture under Linux, we got the code shown in Listing~\ref{lst:yesAsm}. Notice the jump pointer at address {{\tt 0x8152ae7}. IDA Pro was not able to interpret the target addresses and therefore assumed this was an end of a function, identifying {\tt 0x8152aee} and {\tt 0x8152b12} as new functions. It also left code in address range {\tt 0x8152afa} -- {\tt 0x8a52b11} as not within a single function. This error occured even with an unstripped binary with full debug information indicating the correct function start and size.
	
We compiled enscript with clang compiler (Version 6)  using -flto (linktime optimization) and -O2 to get the code in Listing~\ref{lst:enscript} for the function {\tt process\_file}. Here the LTO optimization results in some extra insertion of {\tt nop}s. IDA Pro indicates that the instruction starting at address {\tt 0x24540} is a new function, even though the symbol table includes the correct function boundaries. Also notice the stack adjustment and pushing of the frame pointer in addresses {\tt 0x24540} and {\tt 0x24543}. IDA Pro could have been looking at the long set of {\tt nop}s or this pattern to make its determination. 
	
We know that IDA Pro does take advantage of some of the metadata, because we found less errors when it was run on an unstripped binary versus stripped versions of the same binary, however it still has some issues with function boundary detection even on unstripped binaries. On a test of approximately 57,000 unstripped binaries, in 19\% of the cases, there was not a perfect match between IDA Pro listed function starts and the function starts of the symbol table. In 1\% of the cases, the resulting F1 statistic for correct function starts was less than 96\%. Most of these cases occurred in evaluation of 32-bit binaries. Although this is not a horrible statistic, we want our ground truth to be 100\% correct for validation.

\begin{lstlisting}[caption={Example misindentified function entry in objdump utility (clang 32bit -O2 option), in AT\&T syntax for Intel.}, captionpos=b,label={lst:yesAsm},frame=tb,float]
08152ad0 <get_DW_IDX_name>:
8152ad0:	movl   0x4(%esp),%ecx
8152ad4:	cmpl   $0x1fff,%ecx
8152ada:	jg     8152af4 <get_DW_IDX_name+0x24>
8152adc:	decl   %ecx
8152add:	cmpl   $0x4,%ecx
8152ae0:	ja     8152b30 <get_DW_IDX_name+0x60>
8152ae2:	movl   $0x81f8d07,%eax
8152ae7:	jmpl   *0x81f624c(,%ecx,4)
	--- Misidentified ---
8152aee:	movl   $0x81f8d1b,%eax
8152af3:	retl   
	--- Following Code Orphaned ---
8152afa:	je     8152b24 <get_DW_IDX_name+0x54>
8152afc:	cmpl   $0x2001,%ecx
8152b02:	je     8152b2a <get_DW_IDX_name+0x5a>
8152b04:	cmpl   $0x3fff,%ecx
8152b0a:	jne    8152b30 <get_DW_IDX_name+0x60>
8152b0c:	movl   $0x81f8d5d,%eax
8152b11:	retl   
	--- Misidentified ---
8152b12:	movl   $0x81f8d2c,%eax  
	. . .
\end{lstlisting}

\begin{lstlisting}[caption={Example misindentified function entry (clang 32bit -O2 -flto options), in AT\&T syntax for Intel.}, captionpos=b,label={lst:enscript},frame=tb,float]
  while (fgets (buf, sizeof (buf), ...
24516:	    sub    $0x4,%esp
24519:	    push   %eax
2451a:	    push   $0x1000
2451f:	    lea    0x1274(%esp),%ebp
24526:	    push   %ebp
24527:	    call   34d10 <fgets@plt>
2452c:	    add    $0x10,%esp
2452f:	    test   %eax,%eax
24531:	    je     24691 <process_file+0x4581>
24537: 	    xor    %ebx,%ebx
24539:	    nop
2453a:	    nop
2453b:	    nop
2453c:      nop
2453d:	    nop
2453e:      nop
2453f:	    nop
	--- Misidentified ---
      i = strlen (buf);
24540:	    sub    $0xc,%esp
24543:	    push   %ebp
24544:	    call   34bd0 <strlen@plt>
24549:	    add    $0x10,%esp
. . .
\end{lstlisting}

\begin{lstlisting}[caption={Example misinterpreted function end for function quotearg\_buffer\_restyled (icc 64bit -O0 option)}, captionpos=b,label={lst:badEnd},frame=tb,float,floatplacement=t]
4056b5:     add    %rax,%rdx
4056b8:     mov    (%rdx),%rax
4056bb:  jmpq   *%rax
	--- Misidentified ---
4056bd:     movzbl -0x120(%rbp),%eax
4056c4:     movzbl %al,%eax
4056c7:     test   %eax,%eax
4056c9:     je     40580f
\end{lstlisting}

\begin{lstlisting}[caption={Example missing function entry (clang 64bit -O1), in AT\&T syntax for Intel}, captionpos=b,label={lst:enscriptmissing},frame=tb,float]
408115:    mov    (%rax,%rcx,8),%rcx
408119:    mov    %edx,0x20(%rcx)
40811c:    mov    0x20d35d(%rip),%rcx        
408123:	   mov    %rbx,(%rax,%rcx,8)
408127:    pop    %rbx
408128:    jmpq   407640 
40812d:	   xor    %ecx,%ecx
40812f:	   cmp    %rbx,%rcx
408132:	   jne    4080e5 
408134:	   pop    %rbx
408135:    retq   
408136:	   nopw   %cs:0x0(%rax,%rax,1)
40813d:	00 00 00 

	--- Missing Function ---
0000000000408140 <yyalloc>:
408140:	   jmpq   401450 <malloc@plt>
408145:	   data16 nopw %cs:0x0(%rax,%rax,1)
40814c:	00 00 00 00 
\end{lstlisting}

The code in Listing~\ref{lst:badEnd} is derived from quotearg.c file, function quotearg\_buffer\_restyled. This is part of a standard gnu library for parsing command line arguments that is used by many common utilities. We compiled this library codes as part of a compilation of the coreutils suite of programs, using the intel compiler icc, with -O0 optimization and 64 bit Intel architecture under Linux. For this example, the library is part of the utility test. Here IDA Pro can not determine the target of the jump pointer instruction on line {\tt 0x4056bb} and therefore reports that the instruction on line {\tt 0x4056bd} is the start of a new function.

Any tools that rely on IDA Pro, either as plug-ins or stand alone, should not just rely on IDA Pro when deriving the ground truth from the unstripped binaries.

\subsection{Ghidra}
Ghidra~\cite{Ghidra} is a reverse engineering tool developed by the National Security Agency, and was first publicly released in 2019. Ghidra has many capabilities similar to IDA Pro and can be used for disassembly, detection of function boundaries, etc. As with IDA Pro, it uses available debug and symbol table information. And, as with IDA Pro, it is not perfect in generating ground truth information. We have not seen papers that explicitly use Ghidra for ground truth, but it would not suprise us as more people begin to use it. In general Ghidra has done better, but as seen in Listing~\ref{lst:enscriptmissing}, for the {\tt states} program (which is part of the {\tt enscript} utility}, Ghidra missed some function starts when analyzing the unstripped binary, even though they are listed in the symbol table. Therefore we can not always rely on it either for 100\% accurate ground truth.

Any tools that rely on Ghidra, either as plug-ins or stand alone, should not just rely on Ghidra when deriving the ground truth from the unstripped binaries.

\subsection{Symbol Table}
The symbol table in an unstripped binary contains information about functions and global variables (Listing~\ref{lst:as-new-syms} shows part of a symbol table). The problem we have found when using symbol tables for ground truth are:

\begin{itemize}
	\item They don't contain all of the information we want. For example, no mapping of which bytes are instructions. Jump tables and global pointers do not always have symbols. 
	\item There may not be a one-to-one mapping between function name symbols and source code. The example in Listing~\ref{lst:as-new-syms} shows just that case, where there are multiple binary functions for a specific source function. The example in Listing~\ref{two-entry} shows multiple entry points for the same function. We have also seen aliases for functions, where two names point to the same function location. 
	\item Function lengths are not consistent. We found that the Intel compiler includes padding bytes in the function length in the symbol table, other compilers do not. 
\end{itemize}

\subsection{DWARF}
We have seen several papers state that they use debug data for ground truth, but never elaborate further. When looking at DWARF debug data~\cite{dwarf}, we have seen several issues. First involves correctly interpreting the data. The HIGH PC value in a subprogram is the byte after the end of the function. However, it may be an absolute value or a relative value (length), which is not parsed correctly by some libraries/tools. Not all compilers include complete information in the DWARF data. It is a good source, but researchers have to be careful.

\subsection{Compiler Hacks}
There are a couple of tools that generate ground truth by hacking the compiler. One example of this is the work by Pang et al.~\cite{Pang21}, where the authors revise the compiler to emit all of the ground truth information they need. Li~et~al.~\cite{Li2020} use intermediate representation, such as generated assembly code listings, to assist in the generation of their ground truth for disassembles. 

These techniques only work for the compilers they are designed for, and therefore can not be reliably used for generalization of ground truth, even with newer versions of the same compilers.

\section{Conclusion}
\label{Recommend}
Knowing the ground truth is essential when evaluating the effectiveness of binary analysis tools. We have seen instances where the ground truth was incomplete, misleading, misinterpreted or even hacked to get results that the authors wanted. We are not saying that the authors deliberately misled the community, but rather did not focus on the importance of making sure the ground truth was correct. Most authors do not communicate the details of their generation of ground truth or the assumptions they made when doing the evaluation.

Without the existence of well vetted tools and/or data sets for ground truth, we will struggle with the ability to accurately build, evaluate and gauge binary analysis tools. If researchers then use incorrect ground truth when using machine learning or other automated analysis, the problem will just get worse. We recommend a discussion among the community about the types of ground truth metrics we need, the best ways to develop them, and a process for vetting and sharing ground truth generation tools. 

We do not believe custom tools, such as compilers modifications, are a good long term solution to ground truth generation. Use of DWARF~\cite{dwarf} debug data and the compiler generated symbol tables is a good start, but their limits need to be fully explored. 

Each method of generating a binary, different compilers, architectures, compiler optimizations, and source language creates challenges for binary analysis. We want our tools to be as accurate a possible. Creating good testsuites, with well documented and well create binaries is essential for validating the tools. When creating the testsuites, the researcher has control. They can ensure that all of the symbol table and other information is available in each binary in the testsuite. If, for some reason, it is not possible to have complete ground truth for a particular binary, it should be removed from the testsuite. This is permitted, since the idea of the testsuite is to validate the tools or algorithms, so we need 100\% correct ground truth.
Researchers should check to make sure their ground truth is correct, and they should clearly document how they arrived at the ground truth. Researchers should make scripts and programs for ground truth extraction and measurement of results to that ground truth publicly available.





\bibliographystyle{IEEEtranS}
\bibliography{IEEEabrv,bibfile}

\end{document}